\documentclass[aps,prl,twocolumn,
showpacs,preprintnumbers,amsmath,amssymb]{revtex4}
\usepackage[dvips]{graphicx}
\usepackage{enumerate}

\newcommand{\beq}{\begin{eqnarray}}
\newcommand{\eeq}{\end{eqnarray}}
\newcommand{\non}{\nonumber\\}
\newcommand{\SU}{\text{SU}}
\newcommand{\U}{\text{U}}

\def\6#1{{\underline{#1}}}
\def\m6#1{{\underline{#1}\,}}

\newdimen\Tdim
\def\ispan{{\setbox0=\hbox{i}%
\Tdim\ht0\advance\Tdim\dp0\rule[-\dp0]{0pt}{\Tdim}}}
\def\jspan{{\setbox0=\hbox{j}%
\Tdim\ht0\advance\Tdim\dp0\rule[-\dp0]{0pt}{\Tdim}}}
\def\Tspan#1{{\setbox0=\hbox{#1}%
\Tdim\ht0\advance\Tdim\dp0\advance\Tdim.55ex\rule[-\dp0]{0pt}{\Tdim}\box0}}

\def\Tr{{\rm Tr}}

\def\p{\partial}

\def\D{\mathcal{D}}

\def\=:{=\hspace{-.7em}\raisebox{1.1ex}{.}\hspace{.1em}\raisebox{-0.2ex}{.} }

\begin{document}

\preprint{RIKEN-TH-178, TKYNT-09-23}
\title{Instabilities of Non-Abelian Vortices in Dense QCD}
\author{Minoru Eto$^1$, Muneto Nitta$^2$, and Naoki Yamamoto$^3$}
\affiliation{
$^{1}$Theoretical Physics Laboratory, RIKEN, Saitama 351-0198, Japan\\
$^{2}$Department of Physics, and Research and Education Center for 
Natural Sciences, Keio University, 4-1-1 Hiyoshi, Yokohama, Kanagawa 
223-8521, Japan\\
$^{3}$Department of Physics, The University of Tokyo, Tokyo 113-0033, 
Japan}
\date{\today}

\begin{abstract}
We construct a low-energy effective theory describing non-Abelian 
vortices in the color superconducting quark matter under stress. We 
demonstrate that all the vortices are radically unstable against decay 
into the only one type of vortices due to the potential term induced by 
the explicit flavor symmetry breaking by the strange quark mass. A simple
analytical estimate for the lifetime of unstable vortices is provided
under the controlled weak-coupling calculations. We briefly discuss the 
(non)existence of magnetic monopoles at high density.

\end{abstract}


\maketitle

\emph{Introduction.}---Topological or quantized vortices commonly arise in a wide area of 
physics from condensed matter physics and cosmology to particle 
physics \cite{VS}. In the context of nuclear physics, the dynamical 
breaking of $\U(1)_B$ baryon number due to the neutron superfluidity 
in nuclear matter gives rise to topologically stable vortices 
characterized by the first homotopy group $\pi_1[\U(1)_B] = \mathbb{Z}$. 
They are phenomenologically important since the sudden increase of the 
rotation of neutron stars, the so-called glitches, may be attributed 
to the unpinning of vortices which releases an angular momentum transfer 
from the nuclear ``mantle" to the outer crust \cite{glitch}.
Topological superfluid vortices also emerge in the color superconducting 
quark matter \cite{CSC} presumably existing in the ``core" of neutron 
stars: the $\U(1)_B$ symmetry is broken by the condensation of diquark 
pairs in the color-flavor locked (CFL) phase \cite{ARW99} which is 
indeed shown to be the most stable ground state at asymptotic high 
density in quantum chromodynamics (QCD). Recently, however, it has been 
found that minimal topological vortices in quark matter are not $\U(1)_B$ 
vortices \cite{FZ02} but non-Abelian vortices \cite{vortex} referred to 
as the semisuperfluid vortices \cite{BDM06}, which have only winding 
number $1/3$ inside $\U(1)_B$. Actually it is energetically favorable 
for a single $\U(1)_B$ vortex to split into three (red, green, and blue) 
non-Abelian vortices. At first glance, one may expect that all the 
resultant three non-Abelian vortices are stable. 

In this Letter, we show that these remaining non-Abelian vortices are 
still unstable against decay into the only one type of stable 
vortices when the effect of nonzero strange quark mass $m_s$ is taken 
into account. In order to elucidate the (in)stabilities of non-Abelian 
vortices in a model-independent manner, we use the Ginzburg-Landau (GL) 
approach near the transition temperature $T_c$, and construct the 
low-energy effective theory of vortices with the potential term induced 
by the explicit breaking of flavor symmetry. Owing to the asymptotic 
freedom of QCD, all the calculations throughout this Letter are under 
theoretical control at high density regime where the QCD coupling 
constant is weak. We remark that the existence of non-Abelian vortices 
by itself does not rely on the domain of applicability of the GL 
Lagrangian, but only on the dynamical symmetry breaking induced by the 
diquark condensation. On the other hand, the symmetry argument is not 
enough to ensure their stabilities which depend on the details of the 
dynamics. In the following, we neglect the effect of $\U(1)_{\rm EM}$ 
electromagnetism since the mixing between broken $\SU(3)_C$ color and 
$\U(1)_{\rm EM}$ is sufficiently small at high density. The 
generalization to include the effect is straightforward.

\emph{Ginzburg-Landau Lagrangian.}---We consider the diquark pairing in the most attractive CFL and 
spin-parity $0^+$ channel \cite{ARW99}: 
$(\Phi_L)_a^i \sim \epsilon_{abc} \epsilon_{ijk}
\langle (q_L)_b^j C (q_L)_c^k \rangle$ and 
$(\Phi_R)_a^i \sim \epsilon_{abc} \epsilon_{ijk}
\langle (q_R)_b^j C (q_R)_c^k \rangle$,
where $i,j,k$ ($a,b,c$) are flavor (color) indices and $C$ is the charge 
conjugation operator. Here we take $\Phi_L=-\Phi_R=\Phi$ so that the 
ground state is the positive parity state.

The time-dependent Ginzburg-Landau (TDGL) Lagrangian up to the second 
order in time and space derivatives \cite{TDGL} at large quark 
chemical potential $\mu \gg m_s \gg m_{u,d} \simeq 0$ near $T_c$ is 
given by \cite{GR02, IMTH04}:
\beq
{\cal L}_{\rm GL} &=& 
\Tr\left(K_0\D_0\Phi^\dagger \D_0\Phi 
- K_3\D_i \Phi^\dagger \D_i \Phi \right) \non & & 
+ \Tr\left( K_{\rm D} \Phi^{\dagger}\D_0 \Phi + {\rm H.c.} \right)
- \frac{1}{4} F_{\mu \nu} F^{\mu \nu} - V_{\rm GL}, \non
V_{\rm GL}&=& \Tr\left[\Phi^\dagger\biggl\{
\Bigl(\alpha+\frac{2\epsilon}{3}\Bigr){\bf 1}_3 
+ \epsilon X_3 \biggr\}\Phi\right] \non & & 
+ \beta_1 \left[\Tr(\Phi^\dagger\Phi)\right]^2 
+\beta_2 \Tr \left[(\Phi^\dagger\Phi)^2\right],
\label{eq:gl}
\eeq
where $\D_\mu \Phi = \p_\mu \Phi - i g_s A_\mu \Phi$, 
$F_{\mu \nu}=\partial_{\mu}A_{\nu}
-\partial_{\nu}A_{\mu}-ig_s[A_{\mu},A_{\nu}]$, 
and $X_3 = \frac{1}{2} {\rm diag}(0,1,-1)$.
$K_{\rm D}$ is a dissipative term reflecting the decay of Cooper pairs 
into fermionic excitations.
The $\epsilon$ terms originate from a Fermi surface splitting due to the 
nonzero strange quark mass together with the constraints of electric and 
color charge neutrality and weak interaction equilibration \cite{IMTH04}.

The GL parameters $K_{0,3}$, $\alpha$, $\beta_{1,2}$, and $\epsilon$ 
are obtained from the weak-coupling calculations 
\cite{GR02, IMTH04}:
\beq
& &\! \! \! \! \! \! \! \! \! 
\alpha = 4 N(\mu) \log \frac{T}{T_c},\quad
\beta_{1,2} = \frac{7\zeta(3)}{8(\pi T_c)^2}\, N(\mu)\equiv \beta,\quad 
\non & &\! \! \! \! \! \! \! \! \! 
K_3 = \frac{1}{3}K_0 = \frac{7\zeta(3)}{12(\pi T_c)^2}N(\mu), \ 
\epsilon = N(\mu) \frac{m_s^2}{\mu^2}\log \frac{\mu}{T_c},
\label{eq:weak}
\eeq
where $N(\mu) = {\mu^2}/({2\pi^2})$ is the density of state at the Fermi 
surface and $T_c = 2^{1/3}e^\gamma \Delta/{\pi}$ is the critical 
temperature of the CFL phase in the absence of $m_s$. $K_0$ and $K_{\rm D}$ 
have not been calculated in the literature, but can be derived
following the same procedure of Ref.~\cite{AT66}.

The ground state of the GL potential is given by
$\Phi \! = \! 
\left[ \left(\! -\frac{\alpha}{8 \beta } 
- \frac{\epsilon}{12\beta}\right){\bf 1}_3 
-  \frac{\epsilon}{2\beta}X_3 \right]^{1/2} \! \! \! \! \! 
\equiv \! \! \! {\rm diag}(\Delta_{1},\Delta_{2},\Delta_{3})$,
where the gap parameters $\Delta_1$, $\Delta_2$, and $\Delta_3$ denote
down-strange, strange-up, and up-down Cooper pairs, respectively.
Due to the gap ordering, $\Delta_3>\Delta_1>\Delta_2$,
the symmetry breaking pattern is \cite{IMTH04}
\beq
\label{eq:symmetry}
\SU(3)_C \times \SU(3)_{L,R} \times \U(1)_B
\xrightarrow{\Phi} \SU(3)_{C+F}
\xrightarrow{m_s}  
\U(1)_V^2.
\eeq
For clarity and completeness, we will first neglect the 
$\epsilon X_3$ term and later treat it as a perturbation.
Without the $\epsilon X_3$ term, the order parameter is given 
by $\Phi = \bar\Delta{\bf 1}_3 
\equiv \sqrt{-\frac{\alpha}{8\beta}-\frac{\epsilon}{12\beta}}{\bf 1}_3$.

Mass spectra around this ground state are
\beq
\label{eq:spectra}
m_G^2 \!=\! 2g_s^2 \bar\Delta^2 K_3, \
m_{1}^2 \!=\! - \frac{2}{K_3} \! \! 
\left(\alpha \!+\! \frac{2\epsilon}{3}\right), \
m_{8}^2 \!=\! \frac{4\beta \bar\Delta^2}{K_3},
\eeq
where $m_G$ is the mass of the gluons, $m_1$ and $m_8$ are the masses of 
quarks in the $3 \otimes \bar 3 = 1 \oplus 8$ representation under the 
unbroken $\SU(3)_{C+F}$ symmetry, respectively. From Eqs.~(\ref{eq:weak}) 
and (\ref{eq:spectra}), we have $m_G \sim g_s \mu$ and $m_1 \simeq 2m_8 
\sim \bar \Delta$; then, the relation $g_s \mu \gg \bar \Delta$ at high 
density indicates that the CFL phase is a type-I superconductor 
\cite{GR03}. Note that non-Abelian vortices can appear even in this 
type-I system, since their interactions are repulsive at large distances 
due to the exchange of the Nambu-Goldstone (NG) boson associated with the 
$\U(1)_B$ symmetry breaking \cite{NNM08}. This is in contrast to the case 
of the metallic (Abelian) superconductor where vortices can exist only in 
a type-II system. Non-Abelian vortices are rather superfluid vortices; 
they are created under a rapid rotation.

\emph{Profiles of non-Abelian vortices.}---Corresponding to the three types 
of vortices, the order parameter $\Phi$ asymptotically behaves as
\beq
\Phi \! \xrightarrow{r\to 0} \! \left\{
\begin{array}{l}
{\rm diag}\left(0,*,*\right)\\
{\rm diag}\left(*,0,*\right)\\
{\rm diag}\left(*,*,0\right)
\end{array}
\! \! \right., \ \ 
\Phi \! \xrightarrow{r\to \infty} \! \left\{
\begin{array}{l}
{\rm diag}\left(e^{i\theta} \Delta_{1},\Delta_{2},\Delta_{3}\right)\\
{\rm diag}\left(\Delta_{1},e^{i\theta}\Delta_{2},\Delta_{3}\right)\\
{\rm diag}\left(\Delta_{1},\Delta_{2},e^{i\theta}\Delta_{3}\right)
\end{array}
\right.
\nonumber
\eeq
where $(r,\theta)$ is the polar coordinate and ``$*$'' stand for some 
nonzero constants. All the above three asymptotic forms at infinity can 
be brought into a unique form $\Phi \xrightarrow{r\to \infty} 
e^{i{\theta}/{3}}\ {\rm diag}(\Delta_{1},\Delta_{2},\Delta_{3})$
by regular $\SU(3)_C$ gauge transformations \cite{NNM08}. The overall 
phase $e^{i{\theta}/{3}}$ manifestly shows that the non-Abelian vortex 
winds $2\pi/3$ inside $\U(1)_B$. Their tensions logarithmically diverge 
as $T \simeq \frac{2\pi \bar \Delta^2}{3}\log\frac{L}{r_0} + {\cal O}(1)$ 
where $L$ is a long-distance cutoff and $r_0$ is a short-distance cutoff.

Let us take a diagonal ansatz for a single vortex
\beq
\Phi &=& \bar\Delta\ e^{i\theta\left[\frac{1}{\sqrt3}T_0 
- \sqrt{\frac{2}{3}}\left(\nu_3 T_3 + \nu_8T_8\right)\right]}
\non
& & \times 
\left[\frac{F(r)}{\sqrt3}T_0 - \sqrt{\frac{2}{3}}G(r)
\left(\nu_3 T_3 + \nu_8T_8\right)\right],\\
A_i &=& \frac{1}{g_s}\frac{\epsilon_{ij}x^j}{r^2}[1-h(r)]
\sqrt{\frac{2}{3}}\left(\nu_3 T_3 + \nu_8T_8\right),
\eeq
with $T_0 = \frac{1}{\sqrt 3}{\rm diag}(1,1,1)$, 
$T_3 = \frac{1}{\sqrt 2}{\rm diag}(0,1,-1)$, and
$T_8 = \frac{1}{\sqrt 6}{\rm diag}(-2,1,1)$.
We impose $(F,G,h) \to (3,0,0)$ as $r \rightarrow \infty$ to satisfy 
$\Phi \to \bar\Delta{\bf 1}_3$. The single-valuedness condition for $\Phi$ 
requires $(\nu_3,\nu_8)=(0,1),(\pm\frac{\sqrt{3}}{2},-\frac{1}{2})$. 

In the presence of each vortex, the remaining $\SU(3)_{C+F}$ symmetry is 
further broken down to $[\U(1)\times \SU(2)]_{C+F}$. Hence, the vortex 
solution is labeled by the NG modes (or the orientational modes) living 
on the coset space $\SU(3)_{C+F}/[\U(1) \times \SU(2)]_{C+F} \simeq 
\mathbb{C}P^2$, which we parametrize by introducing 
$\phi=(\phi_1,\phi_2,\phi_3)^T$ 
($\phi^\dagger\phi=1$, $\phi \sim e^{i\alpha} \phi$)
defined as 
$U \left[\sqrt{\frac{2}{3}} \left(\nu_3 T_3 + \nu_8 T_8\right)\right] 
U^\dagger \equiv \phi\phi^\dagger - \frac{1}{3}{\bf 1}_3$.
The most general solution can be obtained by acting 
$U \in \SU(3)_{C+F}$ as $\Phi \to U\Phi U^\dagger$
and $A_i \to UA_{i}U^\dagger$:
\beq
\label{eq:Phi}
\Phi &=& \bar\Delta\, e^{\frac{i\theta}{3}} \left[
\frac{F(r)}{\sqrt 3}T_0 + G(r) 
\left( \phi \phi^\dagger - \frac{1}{3}{\bf 1}_3\right) \right],\\
\label{eq:A}
A_i &=& \frac{1}{g_s} \frac{\epsilon_{ij}x^{j}}{r^2}
h(r) \left( \phi \phi^\dagger - \frac{1}{3}{\bf 1}_3\right).
\eeq

For concreteness, let us choose $(\nu_3,\nu_8)=(0,1)$ as a reference 
solution. Equations of motion for the profile functions read \cite{EN09}:
\beq
& & \! \! \! \! \! \! \! 
f''\! + \! \frac{f'}{r} \!
- \! \frac{(2 h \!+\! 1)^2}{9 r^2}f \! - \! 
\frac{m_{1}^2}{6} f \left(f^2 \!+\! 2 g^2 \!-\! 3\right)
\!-\! \frac{m_{8}^2}{3} \! f \left(f^2 \!-\! g^2\right) \! =\! 0,
\non
& & \! \! \! \! \! \! \! 
g''\!+ \! \frac{g'}{r} \!
- \! \frac{(h \!-\! 1)^2}{9 r^2}g \!- \! 
\frac{m_{1}^2}{6} g \left(f^2 \!+\! 2 g^2\!-\! 3\right) \!
+\! \frac{m_{8}^2}{6} g \left(f^2 \!-\! g^2\right) \!=\! 0,
\non
& & \! \! \! \! \! \! \!
h''-\frac{h'}{r} - \frac{m_G^2}{3}  \left[g^2 (h-1)+f^2 (2 h+1)\right] = 0,
\label{eq:profile}
\eeq
with $f \equiv \frac{1}{3}(F+2G)$ and $g \equiv \frac{1}{3}(F-G)$.
These equations are solved with the boundary conditions,
$(f,g,h) \to (1,1,0)$ as $r \to \infty$ and 
$(f,g',h) \to (0,0,1)$ as $r \to 0$.

\emph{Low-energy effective theory.}---The NG modes $\phi \in \mathbb{C}P^2$ propagate along the non-Abelian 
vortex string. The philosophy of constructing the low-energy effective 
Lagrangian is similar to that of the chiral perturbation theory (ChPT) 
describing the low-energy dynamics of QCD. Remembering that the ChPT is 
constrained by the $[\SU(N_f)_L \times \SU(N_f)_R]/\SU(N_f)$ symmetry, 
the form of the Lagrangian in our case is determined solely by the 
$\SU(3)/[\U(1)\times \SU(2)]$ symmetry, 
and is described by the $\mathbb{C}P^2$ nonlinear sigma model 
\cite{ENN09}:
\beq
{\cal L}_{\mathbb{C}P^2} = C\, \sum_{\alpha=0,3} 
K_\alpha [\p^\alpha\phi^\dagger \p_\alpha \phi 
+ (\phi^\dagger \p^\alpha\phi) (\phi^\dagger \p_\alpha\phi)],
\label{eq:low_lag}
\eeq
where $\phi$ is promoted to a dynamical field as $\phi \to \phi(x_0,x_3)$ 
depending on the vortex world-sheet coordinates $x_0$ and $x_3$.
$K_\alpha$ are the stiffness parameters in Eq.~(\ref{eq:gl}) and we have 
only one unknown constant $C$. Note that the $K_{\rm D}$-term in 
Eq.~(\ref{eq:gl}) gives no contribution to Eq.~(\ref{eq:low_lag}) since
it is traceless in the vortex background solutions \cite{ENN09}.

In order to determine the constant $C$, we have to go back to the 
original GL Lagrangian (\ref{eq:gl}) and we have to know the $\phi$ dependences 
of $\Phi$ and $A_{\mu}$. It is easy for $\Phi$ and $A_{i=1,2}$ because we 
have already solved background vortex solutions as 
$\Phi(x_{1,2};\phi(x_{0,3}))$ and $A_{1,2}(x_{1,2};\phi(x_{0,3}))$.
The missing part is $A_{0,3}(\phi(x_{0,3}))$ which vanishes in the 
background solutions. Therefore we make an ansatz in an appropriate gauge 
following Ref.~\cite{GSY05}:
$A_{\alpha}(x_{1,2};\phi(x_{0,3})) \! = \! \frac{i\rho(r)}{g_s} \! \!
\left[ \phi\phi^\dagger , 
\p_{\alpha}\! \!
\left( \phi\phi^\dagger \right)\right] \  (\alpha \! = \! 0,3)$
where $\rho(r)$ is an unknown function. Then we finally arrive at 
\beq
C &=& \frac{4\pi}{g_s^2} \int dr\ \frac{r}{2}
\left[m_G^2\Bigl( 
(1-\rho)(f-g)^2 + \frac{\rho^2}{2}(f^2+g^2)
\Bigr)  \right. \non
& & \qquad \qquad \qquad
\left. + \frac{(1-\rho)^2 h^2 }{r^2} + \rho'{}^2
\right], 
\label{eq:kahler_class}
\eeq
where $\rho$ should be determined so that the integral 
(\ref{eq:kahler_class}) is minimized. Using the Euler-Lagrange 
equation for $\rho$,
$\rho'' \! + \! \frac{\rho'}{r} \! + \! (1 \!-\! \rho) 
\frac{h^2}{r^2} \! - \! \frac{m_G^2}{2} 
\left[(f^2 \!+\! g^2)\rho \!-\! (f \!-\! g)^2\right] 
\!=\! 0$,
one finds that $C$ is indeed finite and $\phi$ is normalizable \cite{EN09}.

\emph{Unstable non-Abelian vortices.}---We now turn on the $\epsilon X_3$ term and consider the regime 
$\epsilon \ll \alpha$, which allows for an analytical treatment. Since 
this term explicitly breaks $\SU(3)_{C+F}$ symmetry, the NG modes in 
Eq.~(\ref{eq:low_lag}) are lifted via an effective potential over the 
$\mathbb{C}P^2$ space. Let us consider a single vortex whose field 
configuration satisfies Eq.~(\ref{eq:profile}). Variations of its 
tension can be thought of as the potential
\beq
V_{\mathbb{C}P^2} 
= \epsilon \int d^2x\ \Tr\left[\Phi^\dagger X_3 \Phi\right]
= D (|\phi_3|^2 - |\phi_2|^2), \label{eq:pot}
\eeq
where we have used $|\phi_1|^2+|\phi_2|^2+|\phi_3|^2 = 1$ and have 
defined
\beq
\label{eq:D}
D = \pi \epsilon \bar\Delta^2  \int^\infty_0 dr\ r(g^2 - f^2). 
\label{eq:D}
\eeq
Note that $D$ is positive and finite because $g-f$ is always positive 
and gets exponentially small as going away from the vortex \cite{EN09};
thus, the effective potential is well-defined. 

\begin{figure}[t]
\begin{center}
\includegraphics[height=4cm]{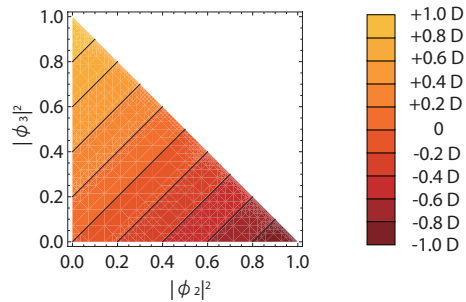}
\end{center}
\vspace{-0.6cm}
\caption{
(Color) Contour plot of the effective potential for the ${\bf C}P^2$ NG 
modes in the $|\phi_2|^2$-$|\phi_3|^2$ plane. The color represents the 
height of the potential.}
\label{fig:pot_cp2}
\end{figure}

The effective potential in $\mathbb{C}P^2$ space is shown in 
Fig.~\ref{fig:pot_cp2}. Since the potential has one minimum at 
$(\phi_1,\phi_2,\phi_3) = (0,1,0)$, any vortices away from $(0,1,0)$ are 
unstable against decay into the $(0,1,0)$ vortex. This matches the fact 
that the pairing gap $\Delta_{2}$ is smaller than $\Delta_{1}$ and 
$\Delta_{3}$ so that the vortex whose string tension is proportional to 
$\Delta_{2}$ is easier to be created than others; the details of the 
dynamics even suggest that the $(1,0,0)$ and $(0,0,1)$ vortices are no 
longer local minima.

Let us estimate the lifetime of unstable vortices. As an example, we 
consider the decay from the $(1,0,0)$ vortex at the left-bottom corner 
of Fig.~\ref{fig:pot_cp2} to the $(0,1,0)$ vortex at the right-bottom corner. 
The discussion here holds for the $(0,0,1)$ vortex. In what follows, we 
set $\phi_3=0$, implying that we will consider a $\mathbb{C}P^1$ 
submanifold (corresponding to the bottom edge of Fig.~\ref{fig:pot_cp2}) 
inside $\mathbb{C}P^2$. It is useful to introduce an inhomogeneous 
coordinate $u(t) \in \mathbb{C}P^1$ by 
$(\phi_1,\phi_2) = ({1}/{\sqrt{1+|u|^2}},{u}/{\sqrt{1+|u|^2}})$.
Then the low-energy effective Lagrangian can be rewritten as
\beq
{\cal L}_{\mathbb{C}P^1} = CK_0 \frac{|\dot u|^2}{(1+|u|^2)^2}
+ {D} \frac{|u|^2}{1+|u|^2}.
\eeq
A typical time scale  of this equation is
$\tau = \sqrt{{CK_0}/{D}}$.

In principle, we can numerically calculate $\tau$ for each $\mu$. 
Here we provide a simple analytical estimate 
instead. Since the profile function $f$ ($g$, $h$, and $\rho$) increases 
(decrease) with a typical scale $r \sim \bar \Delta^{-1}$ for 
$m_G \gg m_{1,8}$ \cite{EN09}, we find $C \sim (\mu/\bar \Delta)^2$ 
from Eq.~(\ref{eq:kahler_class}). Furthermore, $D$ is estimated from 
Eq.~(\ref{eq:D}) as $D \sim \epsilon \sim m_s^2\log({\mu}/{\bar\Delta})$.
Therefore the lifetime of unstable vortices is given by
\beq
\label{eq:lifetime}
\tau \sim {m_s}^{-1} \eta ({\mu}/{\bar\Delta}),\quad
\eta(x) \equiv x^2 \left(\log x\right)^{-1/2}.
\eeq
In the limit $m_s \rightarrow 0$, $\tau \rightarrow \infty$ as anticipated.

\emph{(Non)existence of magnetic monopoles.}---Let us discuss the (non)existence of magnetic monopoles in QCD at high 
density. One may expect that the symmetry breaking pattern 
(\ref{eq:symmetry}) would support the magnetic monopoles characterized 
by $\pi_2[\SU(3)/\U(1)^2]=\mathbb{Z}^2$. If so, monopoles must be 
confined due to the color Meissner effect of the color superconductivity 
because it is in the Higgs phase. In fact, such a confined monopole 
exists in the ${\cal N}=2$ supersymmetric QCD in the Higgs phase with 
the same symmetry breaking pattern (\ref{eq:symmetry}) \cite{SY04}, 
where magnetic fluxes are squeezed into vortex strings confining the 
monopole from both sides. This composite object has been understood 
as a kink in the low-energy effective world-sheet theory on the vortex 
string with a suitable potential term admitting more than or equal to 
two minima. If the low-energy theory (\ref{eq:low_lag}) in our case had a 
potential similar to supersymmetric QCD, this would realize the dual of 
the confinement scenario advocated in the QCD vacuum where monopoles are 
condensed and quarks are confined \cite{N74}. However, this is not the 
case. The potential (\ref{eq:pot}) has only one minimum and allows no 
kink solutions but implies the instabilities of non-Abelian vortices 
instead, as we have seen.

One should note that this conclusion may not be valid if one includes the 
nonperturbative quantum effects which account for the mass gap of NG modes 
as indicated by the Coleman-Mermin-Wagner theorem in two dimensions.
Actually, such effects may lead to multiple local minima in the potential,
and thus, the monopole-antimonopole meson attached to the vortex 
\cite{GSY05, GSY06}.
In the original four-dimensional GL theory at sufficiently high density, 
instanton effects are highly suppressed due to the screening of instantons 
together with the asymptotic freedom of QCD \cite{S01}, and another mechanism 
responsible for the quantum effects should be present.
We will defer this issue to a future work.

\emph{Discussion.}---It is interesting to investigate possible astrophysical implications of 
our results. When the core of a neutron star cools down below the 
critical temperature of the CFL phase, a network of non-Abelian vortices 
will be formed by the Kibble mechanism. Remarkably, the extrapolation of 
our formula (\ref{eq:lifetime}) to the intermediate density regime 
relevant to the core of neutron stars ($\mu \sim 500$ MeV) with 
$\Delta \sim 10$ MeV and $m_s \simeq 150$ MeV suggests that all the 
vortices decay radically with the lifetime of order $\tau \sim 10^{-21}$ 
second. Although this result should be taken with some care due to the 
uncertainty of numerical factor in Eq.~(\ref{eq:lifetime}), it is 
reasonable to expect that only one type of non-Abelian vortices, 
which correspond to the point $(0,1,0)$ in the $\mathbb{C}P^2$ space, 
survive as a response to the rotation of neutron stars in reality. 
The other decaying non-Abelian vortices will emit NG bosons, quarks, 
gluons, or photons during thermal evolution of neutron stars.
In relation to the glitch phenomena, it would be also important to 
understand how the Abelian $\U(1)_B$ vortices in hadronic matter are 
connected to the stable non-Abelian vortices in color superconducting 
quark matter in the interior of neutron stars. This may be relevant to 
the question of continuity of hadronic matter and quark matter 
\cite{SW99,HTYB06}.

M.E. is supported by Special Postdoctoral Researchers Program at RIKEN. 
M.N. is supported in part by Grant-in-Aid for Scientific Research 
(No. 20740141) from the Ministry of Education, Culture, Sports, Science 
and Technology-Japan.
N.Y. is supported by the Japan Society for the Promotion of Science for 
Young Scientists. 
 


\begin{thebibliography}{99}
\bibitem{VS}
A. Vilenkin and E. P. S. Shellard,
{\it Cosmic Strings and Other Topological Defects}
(Cambridge University Press, Cambridge, UK, 1994).

\bibitem{glitch}
P. W. Anderson and N. Itoh,
Nature {\bf 256} 25 (1975).

\bibitem{CSC} 
M. G. Alford {\it et al.},
Rev. Mod. Phys. {\bf 80}, 1455 (2008).

\bibitem{ARW99} 
M. G. Alford, K. Rajagopal, and F. Wilczek, 
Nucl. Phys. {\bf B537}, 443 (1999).

\bibitem{FZ02}
M. M. Forbes and A. R. Zhitnitsky, Phys. Rev. D {\bf 65}, 
085009 (2002); 
K. Iida and G. Baym, Phys. Rev. D {\bf 66}, 014015 (2002).

\bibitem{vortex}
Non-Abelian vortices due to the color-flavor locking were first found 
in the ${\cal N}=2$ supersymmetric QCD:
A.~Hanany and D.~Tong,
J. High Energy Phys. 0307 (2003) 037;
R.~Auzzi {\it et al.}, 
Nucl.\ Phys.\ {\bf B673}, 187 (2003).

\bibitem{BDM06}
A. P. Balachandran, S. Digal, and T. Matsuura,
Phys. Rev. D {\bf 73}, 074009 (2006).

\bibitem{TDGL}
The TDGL Lagrangian (\ref{eq:gl}) is valid near $T_c$ in the 
long-wavelength and low-frequency $\omega$ region satisfying 
$\Delta \ll \omega$ with the BCS gap $\Delta$.

\bibitem{GR02}
I. Giannakis and H.-c. Ren,
Phys. Rev. D {\bf 65}, 054017 (2002);
K. Iida and G. Baym,
Phys. Rev. D {\bf 63}, 074018 (2001); {\bf 66}, 059903(E) (2002). 

\bibitem{IMTH04}
K. Iida {\it et al.}, 
Phys. Rev. Lett. {\bf 93}, 132001 (2004).

\bibitem{AT66}
E. Abrahams and T. Tsuneto, Phys. Rev. {\bf 152}, 416 (1966);
C. A. R. S\'a de Melo, M. Randeria, and J. R. Engelbrecht, 
Phys. Rev. Lett. {\bf 71}, 3202 (1993).

\bibitem{GR03}
I. Giannakis and H.-c. Ren,
Nucl. Phys. {\bf B669}, 462 (2003). 

\bibitem{NNM08}
E. Nakano, M. Nitta, and T. Matsuura,
Phys. Rev. D {\bf 78}, 045002 (2008);
Prog.\ Theor.\ Phys.\ Suppl.\  {\bf 174}, 254 (2008).

\bibitem{EN09}
M. Eto and M. Nitta,
Phys. Rev. D {\bf 80}, 125007 (2009).

\bibitem{ENN09}
M. Eto, E. Nakano, and M. Nitta,
Phys. Rev. D {\bf 80}, 125011 (2009).

\bibitem{GSY05}
A. Gorsky, M. Shifman, and A. Yung,
Phys. Rev. D {\bf 71}, 045010 (2005).

\bibitem{SY04}
M. Shifman and A. Yung, Phys. Rev. D {\bf 70}, 045004 (2004);
D. Tong, Phys. Rev. D {\bf 69}, 065003 (2004);
A. Hanany and D. Tong, J. High Energy Phys. 04 (2004) 066;
M.~Eto {\it et al.}, 
Phys.\ Rev.\  D {\bf 72}, 025011 (2005);
J.\ Phys.\ A  {\bf 39}, R315 (2006).
  
\bibitem{N74}
  Y.~Nambu, Phys.\ Rev.\  D {\bf 10}, 4262 (1974);
  G.~'t Hooft, Nucl.\ Phys.\ {\bf B190}, 455 (1981);
  S.~Mandelstam, Phys.\ Rept.\  {\bf 23}, 245 (1976).

\bibitem{GSY06}
  A.~Gorsky, M.~Shifman, and A.~Yung,
  Phys.\ Rev.\  D {\bf 73}, 065011 (2006);
  M.~Shifman and A.~Yung, Rev. Mod. Phys. {\bf 79}, 1139 (2007);
  an expanded version in Cambridge University Press, 2009.

\bibitem{S01}
T. Sch\"{a}fer, Phys. Rev. D {\bf 65}, 094033 (2002);
N. Yamamoto, J. High Energy Phys. 12 (2008) 060. 

\bibitem{SW99}
T. Sch\"{a}fer and F. Wilczek,
Phys. Rev. Lett. {\bf 82}, 3956 (1999). 

\bibitem{HTYB06}
T. Hatsuda {\it et al.}, 
Phys. Rev. Lett. {\bf 97}, 122001 (2006);
N. Yamamoto {\it et al.},
Phys. Rev. D {\bf 76}, 074001 (2007).

\end{thebibliography}
\end{document}